\documentclass[aps,pra,reprint,groupedaddress,amsmath,amssymb]{revtex4-1}
\pdfoutput=1
\usepackage{graphicx}
\usepackage{dcolumn}
\usepackage{bm}
\usepackage{mathtools}

\begin{document}

\title{Propagation of intense few-cycle laser pulse in a telecommunication-type optical fiber: analysis of the error introduced by the unidirectional approximation}

\author{Leonid Konev}
\email{leonides.konev@gmail.com}
\affiliation{Department of Photonics and Optical Information, ITMO University, Saint-Petersburg, Russia}

\author{Yuri Shpolyanskiy}
\affiliation{Department of Photonics and Optical Information, ITMO University, Saint-Petersburg, Russia}

\date{\today}

\begin{abstract}
Propagation of few-cycle laser pulses in optical fiber is investigated beyond the unidirectional 
approximation. Considered medium parameters include dispersion, Kerr and Raman nonlinearities. High 
intensities of $\sim 10^{10}\ \textrm{kW}/\textrm{cm}^{2}$ lead to generation of the octave-spanning
continuum. The error of the unidirectional approximation is seen in simulations as two pulses 
propagating forwards and backwards, both remain weak under the considered conditions. Direct 
analytical estimate of the error is derived and verified numerically. It can be applied to a priory 
justification of the unidirectional approach.
\end{abstract}


\maketitle

\section{Introduction}

Theoretical analysis of the evolution of intense few-cycle laser pulses in optical fibers has been a
topic of interest in modern optics for the past decades. One of the most frequently used approaches
to the simulation of this process relies on the unidirectional approximation, which implies 
propagation of electromagnetic fields in the forward direction along the waveguide axis 
\cite{agrawal:01,kolesik:04}. This reduction of Maxwell’s equations fully ignores fields that 
propagate backwards. If intensities are low and nonlinearities do not play a role, ``forward'' and 
``backward'' fields run independently, and the approximation does not introduce any significant 
error for the homogeneous waveguide. At higher intensities, nonlinear response of the medium becomes
more pronounced and coupling between forward and backward waves occurs. In recent papers, 
\cite{Kinsler:05,rosanov:08,rosanov:09} the importance of this coupling is addressed. Despite 
unavoidable neglect of the backward wave in the unidirectional approach, it provided proper 
descriptions for the large array of experimental results \cite{bespalov:02,agrawal:01,akhmanov:88,brabec:00}. 
Several papers are dedicated to the analysis of the limitations of the unidirectional equations 
\cite{Kinsler:05,rosanov:08,rosanov:09}. However the resulting conclusions do not seem to allow for 
a priori estimations of adequacy of the approximation.

In this paper we attempt to cover this gap. We examine, numerically and analytically, the evolution 
of the fields neglected in the unidirectional approach which is applied to the propagation of 
intense few-cycle laser pulses in the optical fiber with the non-resonant dispersion and the cubic 
nonlinearity including electronic and Raman mechanisms. We solve the set of equations for the 
interaction of the counter-propagating waves and obtain the initial field distributions matched with
the nonlinear response of the medium \cite{Kinsler:05,konev:14}. Profiles of forward and backward 
waves are presented for the normal and anomalous dispersion including the regime of significant 
self-steepening. Without the mentioned matching a reflected wave appears due to the form of the 
boundary conditions \cite{Kinsler:05,konev:14}. A simple analytical estimate of its 
amplitude is obtained. It shows that the reflected wave is weak in transparent media without 
interfaces. The amplitude is proportional to the amplitude of the forward wave and has the order of 
magnitude equal to the nonlinear refractive index divided by the linear refractive index. The 
equation for backward wave produces an additional field that propagates forward together with the 
forward wave, which is always neglected in the unidirectional approximation. It has the same order 
of amplitude as the non-matched reflected wave.

\section{Equations}

We consider the propagation of linearly polarized few-cycle femtosecond pulses in a single-mode 
telecommunication-type fused silica fiber. The presented theoretical approach is not restricted to 
this specific case and can be applied to other transparent optical media with non-resonant 
dispersion and nonlinearity. Transverse field distribution defined by the mode structure is assumed 
to be invariable during the pulse propagation. It allows for reduction of Maxwell’s equations in 
non-magnetic optical waveguide to the scalar second-order wave equation for the spectrum of 
electrical field, which can be written in the following form  \cite{Kinsler:05,boyd:08}:
\begin{equation}
\partial_{z}^{2} G = -k^{2}(\omega) G - k^{2}(\omega) N_{\omega}(E),
\label{eq:wave_equation}
\end{equation}
\begin{equation}
N_{\omega} = \dfrac{4 \pi F[P_{\textrm{NL}}(E)]}{n^{2}(\omega)},
\label{eq:nl_operator_freq}
\end{equation}
where $N_{\omega}(E)$ is the spectral representation of the nonlinear operator applied to the field;
$P_{\textrm{NL}}(E)$ is the nonlinear medium response; $G = F[E]$ is the field spectrum; 
$k(\omega) = \omega n(\omega) / c$ is the propagation constant; $n(\omega)$ is the refractive index;
$\omega$ is the frequency; $c$ is the velocity of light in vacuum; $z$ is the propagation coordinate
along waveguide axis; $t$ is the time coordinate; $F$ denotes the Fourier transform:
\begin{equation}
G(z, \omega) = F[E(z, t)] = \int_{-\infty}^{+\infty} E(z, t) \exp(i \omega t) \textrm{d}t.
\label{eq:fourier}
\end{equation}

As it was shown by a number of authors \cite{Kinsler:05,Kinsler:10,Ferrando:05}, one can easily 
factorize equation (\ref{eq:wave_equation}) after transformation into the wave vector space (see 
e.g. Appendix A in \cite{Kinsler:10}). If the space frequency is denoted as $\eta$, this yields two 
terms: $\eta - k$ and $\eta + k$, corresponding to the ``forward'' ($E_{+}$) and ``backward'' 
($E_{-}$) waves, respectively:
\begin{equation}
E = E_{+} + E_{-},
\label{eq:total_field}
\end{equation}
\begin{equation}
\begin{dcases}
\partial_{z} G_{+} = + ik(\omega)G_{+} + \dfrac{1}{2} ik(\omega) N_{\omega}(E_{+} + E_{-}),\\
\partial_{z} G_{-} = - ik(\omega)G_{-} - \dfrac{1}{2} ik(\omega) N_{\omega}(E_{+} + E_{-}).
\label{eq:direcional_fields_set}
\end{dcases}
\end{equation}

By differentiating with respect to $z$ and summing up Eqs. (\ref{eq:direcional_fields_set}) it is 
easy to see that the total field (\ref{eq:total_field}) satisfies (\ref{eq:wave_equation}).

If $E_{-}$ is ignored in (\ref{eq:direcional_fields_set}), only the first equation for $E_{+}$ 
remains, which is essentially the same as the unidirectional pulse propagation equation 
\cite{tani:14,kolesik:04}.

\section{Medium response}

Let us now complete the set of equations (\ref{eq:direcional_fields_set}) with a detailed model of 
the medium response. The necessary ingredients in the field of intense few-cycle pulses are the 
linear dispersion and the nonlinearity of the medium. As we consider propagation distances less than
$2\ \textrm{mm}$, the linear absorption of fused silica is insignificant and can be ignored.

Non-resonant dispersion of the refractive index of fused silica in its range of transparency can be 
described with high accuracy by the Sellmeier's formula \cite{agrawal:01}:
\begin{equation}
n(\omega)^{2} = 1 + \sum_{i} \dfrac{B_{i} \omega_{i}^{2}}{\omega_{i}^{2} - \omega^{2}},
\label{eq:sellmeier}
\end{equation}
where $i = \{1, 2, 3\}$; $B_{1} = 0.6961663$, $B_{2} = 0.4079426$ and $B_{3} = 0.8974794$ are the 
amplitudes of the resonances; $\omega$ is the frequency of the field; 
$\omega_{i} = 2 \pi c / \lambda_{i}$ are the resonant frequencies, 
$\lambda_{1} = 0.0684043\ \mu m$, $\lambda_{2} = 0.1162414\ \mu m$, $\lambda_{3} = 9.896161\ \mu m$ 
are the resonant wavelengths of the medium.

The two dominant components for the nonlinear interaction of the fused silica with the field of 
intense ultrashort pulses are the instantaneous electronic cubic nonlinearity and the low-inertial 
cubic Raman nonlinearity. The standard phenomenological model described in 
\cite{platonenko:66,jalali:06} incorporates both of them:
\begin{eqnarray}
\begin{cases}
P_{\textrm{NL}} = \chi_{3,\textrm{e}} E^{3} + \chi_{3, \textrm{ev}} RE,  \\
\partial_{t}^{2} R + \dfrac{1}{T_{\textrm{v}}} \partial_{t} R + \omega_{\textrm{v}}^{2} R = \gamma E^{2}.
\end{cases}
\label{eq:nonlinearity}
\end{eqnarray}
Here $R$ is the amplitude of the molecular oscillations, which describes connection between the 
electrical field of the pulse and the oscillation of the molecules in the medium; $T_{\textrm{v}}$ 
is the relaxation time; $\omega_{\textrm{v}}$ is the vibrational frequency of molecules; 
$\chi_{3,\textrm{e}}$ and $\chi_{3,\textrm{ev}}$ are the Kerr and Raman cubic nonlinear 
susceptibilities, respectively; $\gamma$ is the coefficient that characterizes the relation between 
the Kerr and Raman responses. In (\ref{eq:nonlinearity}), $R$ and its first time derivative are zero
before the pulse appears. The nonlinear refractive index coefficient $n_{2}$ in CGS units can then 
be expressed as the sum of components related to $\chi_{3,\textrm{e}}$ and $\chi_{3,\textrm{ev}}$ 
respectively as \cite{akhmanov:88,kozlov:99}:
\begin{equation}
n_{2} = n_{2,\textrm{e}} + n_{2,\textrm{ev}},\ n_{2,\textrm{e}} = \dfrac{3 \pi \chi_{3,\textrm{e}}}{n(\omega_{0})},\ n_{2,\textrm{ev}} = \dfrac{2 \pi \gamma \chi_{3,\textrm{ev}}}{\omega_{\textrm{v}}^2 n(\omega_{0})}.
\label{eq:nl_coefficients}
\end{equation}

For fused silica, nonlinear refractive indices are estimated to have values 
$\tilde{n}_{2,\textrm{e}} = 2.9 \times 10^{-20} m^{2}/W$ and 
$\tilde{n}_{2,\textrm{ev}} = 7.25 \times 10^{-21} m^{2}/W$ in SI units \cite{agrawal:01,kozlov:99}.
Here, $\omega_0$ is the central frequency of radiation spectrum.

\section{Initial field distribution}

The method we use requires the knowledge of the full time profile of the field at a given point on 
the $z$ axis. This requirement is natural for the so-called $z$-propagated equations, such as the 
set (\ref{eq:direcional_fields_set}). Nonlinear and dispersive effects are conveniently covered by 
this type of equations. While time spectra can be observed directly in experiments, it is essential 
that these equations describe their evolution. Another important aspect is that the dispersion 
response of transparent media is represented naturally (and in the simplest form!) in the frequency 
domain. It makes the approach very productive and provides the explanation for experimental results 
\cite{bespalov:02,agrawal:01,akhmanov:88,brabec:00}. However, the process of injection of a given 
time profile into the fiber is not addressed, which introduces difficulties with physical 
interpretation of the problem setup \cite{rosanov:16}. Generically, both forward and backward waves 
have to be known at the input and output sections of the waveguide to form a certain profile. The 
forward wave is assumed to be given at the input, but the backward wave is usually unknown there, 
which breaks the symmetry. This issue is immanent for $z$-propagated equations and cannot be 
avoided. However, in the case of the unidirectional approximation this limitation is less pronounced
because it is shaded by a stronger assumption of insignificance of the backward wave. If we consider
equations for the forward and backward waves and set the initial distribution of the latter to be 
zero at $z = 0$, the problem formulation becomes similar to that for the unidirectional 
approximation. It allows us to reveal the difference between the second-order wave equation 
(\ref{eq:wave_equation}) written in form of (\ref{eq:direcional_fields_set}) and the unidirectional 
approximation.

As it was shown in our previous paper \cite{konev:14}, a backward-propagating field appears
in the beginning of the simulation in such a setup. The calculated backward wave consists of two 
sub-pulses propagating in opposite directions. The part propagating backward $E_{-}^{\textrm{left}}$
does not change its spectral form after its formation and separation from the initial pulse. The 
other part $E_{-}^{\textrm{right}}$ is tightly coupled with the forward wave, its evolution is 
induced by the forward wave during the whole simulation process.

We examine this behavior by looking at the simplified set of equations for the forward and backward 
waves, that describes their propagation over the distances of just a few wavelengths. The 
forward wave does not evolve over distances that are so short, it only propagates with the group 
velocity. Therefore, higher order dispersion terms can be ignored as well as the nonlinearity. As 
for the backward wave, its presence in the considered setup comes from the nonlinear term 
of the original set, which therefore has to be included in the resulting equation together with the 
backward move with the group velocity. It gives the following simplified set of equations:
\begin{eqnarray}
\begin{cases}
\partial_z E_{+} = - \dfrac{1}{V_{\textrm{g}}} \partial_t E_{+}, \\
\partial_z E_{-} = + \dfrac{1}{V_{\textrm{g}}} \partial_t E_{-} + \dfrac{2 \pi}{cn(\omega_{0})} \partial_t P_{\textrm{NL}}(E_{+}),
\end{cases}
\label{eq:simplified_set}
\end{eqnarray}
where $V_{\textrm{g}} = n_{\textrm{g}} / c$ is the group velocity and $n_{\textrm{g}}$ is the group 
refractive index, both calculated at $\omega_{0}$. The forward wave in this case is just the 
solution of the transport equation:
\begin{equation}
E_{+}(z, t) = E_{+}(0, t - z / V_{\textrm{g}}) = E_{0+}(t - z / V_{\textrm{g}}),
\label{eq:forward_simplified}
\end{equation}
where $E_{0+}$ is the initial distribution of the forward wave. The backward wave can also be 
written out analytically as the sum of parts propagating in negative (left) and positive (right) 
directions of the $z$ axis:
\begin{equation}
E_{-}(z, t) = E_{-}^{\textrm{left}}(z, t) + E_{-}^{\textrm{right}}(z, t),
\label{eq:backward_simplified_sum}
\end{equation}
\begin{equation}
\begin{aligned}
E_{-}^{\textrm{left}}(z, t) &= E_{0-}(t + z / V_{\textrm{g}}) + \\
&+ \dfrac{\pi}{n_{0} n_{\textrm{g}}} P_{\textrm{NL}}[E_{0+}(t + z / V_{\textrm{g}})],
\end{aligned}
\label{eq:backward_simplified_left}
\end{equation}
\begin{equation}
E_{-}^{\textrm{right}}(z, t) = - \dfrac{\pi}{n_{0} n_{\textrm{g}}} P_{\textrm{NL}}[E_{+}(z, t)],
\label{eq:backward_simplified_right}
\end{equation}
where $n_{0} = n(\omega_{0})$.

The presence of the forward-running part (\ref{eq:backward_simplified_right}) in the backward wave 
looks surprising at first, but it can be justified both analytically and numerically. From 
(\ref{eq:backward_simplified_right}) we see that this part is fully induced by the forward wave, 
runs along with it and has the opposite sign. The backward-running part 
(\ref{eq:backward_simplified_left}) includes two terms. The first term represents the initial 
distribution of the backward wave moving with the group velocity in the negative $z$ direction. The 
second one is induced by the forward wave. These terms eliminate each other if we take the initial 
distribution of the form:
\begin{equation}
E_{0-}(t) = - \dfrac{\pi}{n_0 n_\textrm{g}} P_{\textrm{NL}}[E_{0+}(t)].
\label{eq:backward_initial}
\end{equation}
This is the so-called matched distribution \cite{Kinsler:05}. Note that the 
nonlinear response appears in (\ref{eq:backward_simplified_left}), 
(\ref{eq:backward_simplified_right}) without concretization of its form. For the nonlinear response 
(\ref{eq:nonlinearity}) the analytical estimate of the amplitude of the backward wave compared to 
the amplitude of the forward wave is:
\begin{equation}
\Gamma = \left| \dfrac{E_{-}^{\textrm{right}}}{E_{+}} \right| = \left| \dfrac{\pi P_{\textrm{NL}}(E_{0+})}{n_0 n_{\textrm{g}}E_{0+}} \right| = \left| \dfrac{n_{2,\textrm{e}}E_{0+}^{2}}{3n_{\textrm{g}}} + \dfrac{\omega_{\textrm{v}}^{2} n_{2,\textrm{ev}} R(E_{0+})}{2 \gamma n_{\textrm{g}}} \right|.
\label{eq:estimate}
\end{equation}

$R(E_{0+})$ can be obtained analytically or numerically from the second equation in 
(\ref{eq:nonlinearity}). Note, that (\ref{eq:estimate}) depends only on the properties of the medium
and the initial distribution of the forward wave, hence it can be used for the a priori estimation 
of the expected error originating from the assumption of the unidirectional propagation.

\section{Simulation results}

We solve system (\ref{eq:direcional_fields_set}) numerically with Fourier split-step method 
\cite{agrawal:01}. Each step $\Delta z$ is split into 2 sub-steps. First nonlinearity is accounted 
for in the time domain using the Crank-Nicolson scheme with internal iterations until convergence is
reached. Next, the dispersive effects are calculated in the frequency domain. The error is analyzed 
by doubling the number of steps in $z$ and $t$. The conservation of the total energy is checked as 
well.

For the model of the initial distribution of a few-cycle forward pulse we take Gaussian profile in 
the form:
\begin{equation}
E_{0+}(t) = E_{\textrm{max}} \exp(-2t^{2} / T_{0}^{2}) \sin(\omega_0 t),
\label{eq:forward_initial}
\end{equation}
where $E_{\textrm{max}}$ is the amplitude of the pulse; 
$\omega_{0} = 2 \pi c / \lambda_{0}$, $\lambda_{0}$ is the central wavelength; 
$\Delta \tau_{0} = NT_{0}$ is the duration; $T_{0} = \lambda_{0} / c$ is the period; $N$ is the 
number of oscillations and $I$ is the peak intensity.

First we simulate the propagation of the pulse under the conditions of normal group dispersion. The 
pulse parameters are 
$\lambda_{0} = 780\ \textrm{nm}$, $\Delta \tau_{0} = 6T_{0} = 15.6\ \textrm{fs}$ and 
$I = 2 \times 10^{10}\ \textrm{kW} / \textrm{cm}^{2} ( I [\textrm{kW} / \textrm{cm}^{2}] = (3n(\omega_{0}) / 8\pi) (E_{\textrm{max}})^{2} [CGS]$
\cite{akhmanov:88}). The backward wave is matched to the nonlinear response of the medium at $z = 0$
using (\ref{eq:backward_initial}) in order to avoid artificial self-reflection.

\begin{figure}[htbp]
\centering
\fbox{\includegraphics[width=\linewidth]{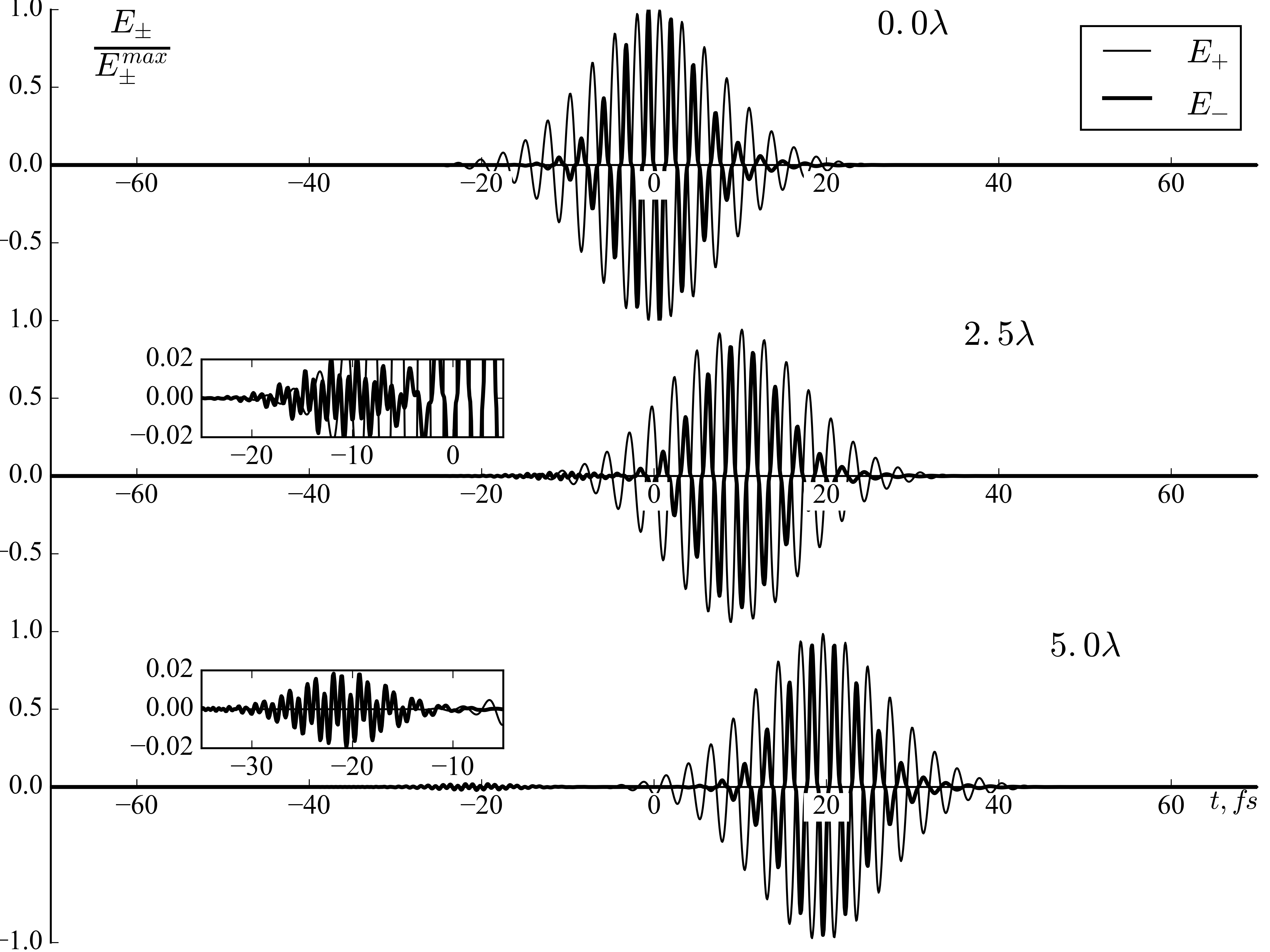}}
\caption{Normalized field profiles of the forward $E_{+}$ and backward $E_{-}$ waves at the initial 
stage of the propagation. Input parameters of the forward wave: central wavelength 
$\lambda_{0} = 780\ \textrm{nm}$, duration $\Delta \tau_{0} = 15.6\ \textrm{fs}$ and intensity 
$I = 2 \times 10^{10}\ \textrm{kW}/\textrm{cm}^{2}$. The backward wave at $z = 0$ is taken in the 
form of (\ref{eq:backward_initial}) to be matched with the nonlinear response of the medium. Ratio 
of the field amplitudes $E_{-}^{\textrm{max}} / E_{+}^{\textrm{max}} \approx 2.7 \times 10^{-3}$. 
The part of the backward wave propagating backwards is given scaled-up in insets.}
\label{fig:normal_initial_field}
\end{figure}

The pulse propagation over the first 5 wavelengths is visualized in Fig. 
\ref{fig:normal_initial_field}. The forward and backward waves are depicted by the thin and thick 
curves, respectively. One can see both the $E_{-}^{\textrm{left}}$ and $E_{-}^{\textrm{right}}$ 
parts of the backward wave. The part $E_{-}^{\textrm{right}}$ that propagates to the right along 
with the forward pulse agrees very well with (\ref{eq:backward_simplified_right}) obtained from the 
simplified system (\ref{eq:simplified_set}); the difference is indistinguishable on the plot. The 
amplitude of this numerically calculated field is approximately 
$2.7 \times 10^{-3} E_{\textrm{max}}$, which agress with the estimate (\ref{eq:estimate}). The part 
$E_{-}^{\textrm{left}}$ is $\sim 25$ times weaker than $E_{-}^{\textrm{right}}$, its profiles are 
given scaled-up in insets. At $z = 0$ $E_{-}^{\textrm{left}}$ is zero, but it is fully formed over 
the first 5 wavelengths and moves to the left. The simplified system (\ref{eq:simplified_set}) 
predicts the absence of $E_{-}^{\textrm{left}}$ for the matched conditions. However, in the system 
(\ref{eq:direcional_fields_set}) higher order dispersion terms originating from (\ref{eq:sellmeier})
are significant, hence the full matching requires consideration of their contribution to the 
refractive index. They are ignored in (\ref{eq:simplified_set}), and we attribute the very weak, but
non-zero $E_{-}^{\textrm{left}}$ to the neglect of dispersion effects in derivation of the matching 
condition (\ref{eq:backward_initial}). It is confirmed by the fact that $E_{-}^{\textrm{left}}$ does
not grow after the initial generation and its spectral density does not vary as will be shown below.

\begin{figure}[htbp]
\centering
\fbox{\includegraphics[width=\linewidth]{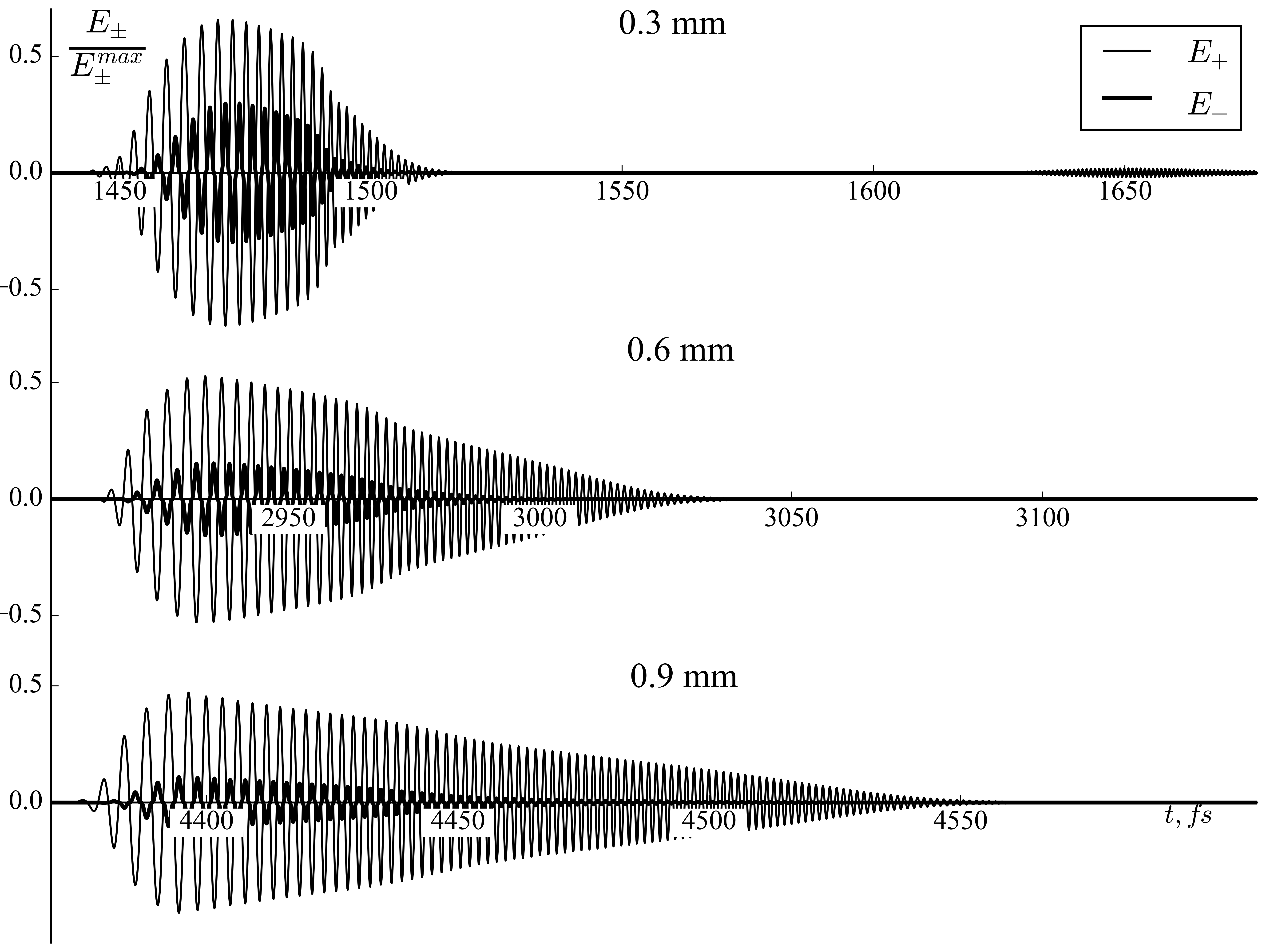}}
\caption{Normalized field profiles of the forward and backward waves after propagation over the 
distances of $z = 0.3,\ 0.6$ and $0.9\ \textrm{mm}$. Parameters are identical to those in Fig. 
\ref{fig:normal_initial_field}.}
\label{fig:normal_slices_field}
\end{figure}

\begin{figure}[htbp]
\centering
\fbox{\includegraphics[width=\linewidth]{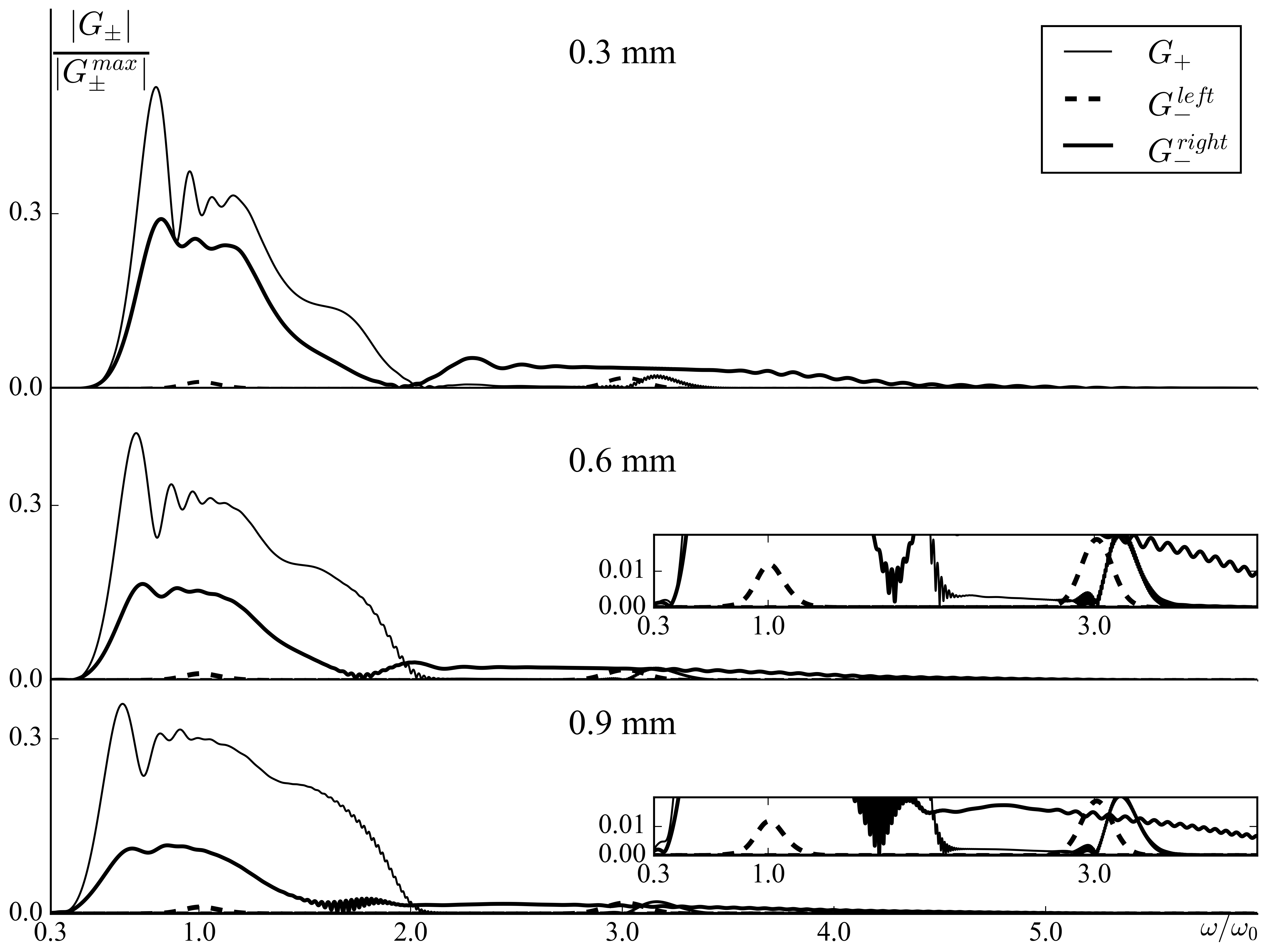}}
\caption{Normalized spectral densities of the fields of forward and backward waves after propagation
over the distances of $z = 0.3,\ 0.6$ and $0.9\ \textrm{mm}$. Parameters are identical to those in 
Fig. \ref{fig:normal_initial_field}.}
\label{fig:normal_slices_spectrum}
\end{figure}

Distributions of $E_{+}$ and $E_{-}$ at further distances $z$ up to $0.9\ \textrm{mm}$ can be seen 
in Fig. \ref{fig:normal_slices_field}. Limits of the $t$ axis in the plots are adjusted to depict 
fields propagating forwards, so that $E_{-}=E_{-}^{\textrm{right}}$ in those $(z,t)$ regions. The 
forward pulse exhibits considerable broadening and shaping under the combined effect of the 
nonlinearity and dispersion with appearance of a chirp typical for the normal dispersion: higher 
frequencies appear at the tail of the pulse. Clearly, the numerical solution does not satisfy the 
simplified system (\ref{eq:simplified_set}). However, the field $E_{-}^{\textrm{right}}$ is not only
fully defined by the forward field $E_{+}$, but, surprisingly, remains related to the latter with 
(\ref{eq:backward_simplified_right}). The field $E_{-}^{\textrm{left}}$ running to the left is 
beyond the visualized $(z,t)$ regions. Formed by $z = 5\lambda_{0}$ (see insets in Fig. 
\ref{fig:normal_initial_field}), this field changes slowly under the influence of the dispersion. 
Normalized spectral densities $G_{+}$, $G_{-}^{\textrm{right}}$, and $G_{-}^{\textrm{left}}$ of 
$E_{+}$, $E_{-}^{\textrm{right}}$, and $E_{-}^{\textrm{left}}$, respectively, are shown in Fig. 
\ref{fig:normal_slices_spectrum}. It can be seen that $E_{-}^{\textrm{left}}$ does not vary from 
$z = 0.3\ \textrm{mm}$ to $z = 0.9\ \textrm{mm}$, i.e. it is not affected by the nonlinearity.
Furthermore the spectrum of the forward pulse is super-broadened, its main part is asymmetric 
ranging from $\sim 0.4\omega_{0}$ to $\sim 2\omega_{0}$ and includes triple frequencies that are 
inherent to the cubic nonlinearity. The spectrum $G_{-}^{\textrm{right}}$ is even broader due to 
(\ref{eq:backward_simplified_right}), it reaches the fifth harmonic.

\begin{figure}[htbp]
\centering
\fbox{\includegraphics[width=\linewidth]{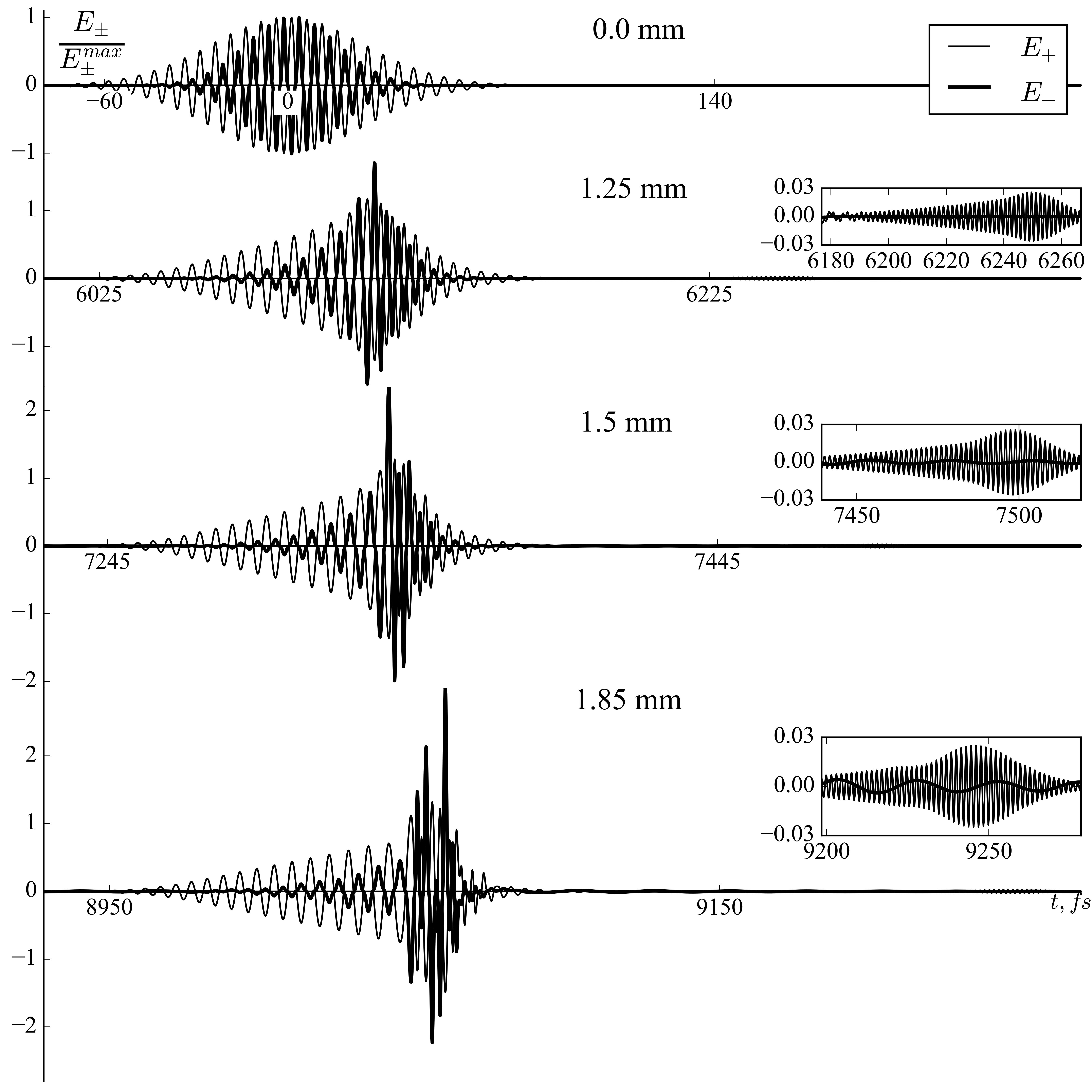}}
\caption{Field distributions of the forward and backward waves of the pulse with intensity 
$I = 5 \times 10^{9} \textrm{kW} / \textrm{cm}^{2}$ after propagation over the distance of 
$z = 0.0,\ 1.25,\ 1.5$ and $1.85\ \textrm{mm}$ under conditions of anomalous dispersion, 
$E_{-}^{\textrm{max}} / E_{+}^{\textrm{max}} \approx 0.8 \times 10^{-3}$.}
\label{fig:anomalous_slices_field}
\end{figure}

\begin{figure}[htbp]
\centering
\fbox{\includegraphics[width=\linewidth]{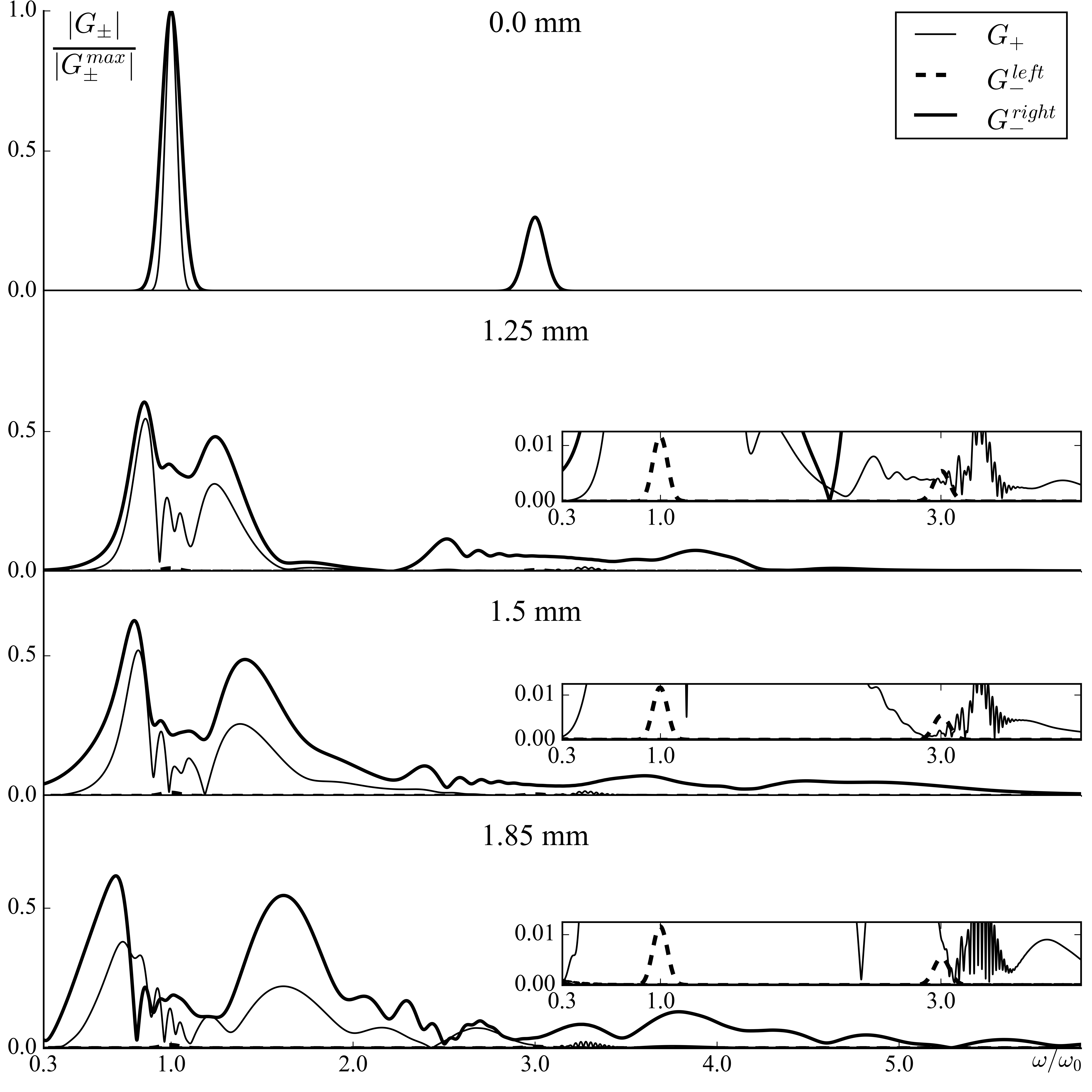}}
\caption{Normalized spectral densities of the fields of forward and backward waves after propagation
over the distances of $z = 0.0,\ 1.25,\ 1.5$ and $1.85\ \textrm{mm}$. Parameters are identical to 
those in Fig. \ref{fig:anomalous_slices_field}.}
\label{fig:anomalous_slices_spectrum}
\end{figure}

\begin{figure}[htbp]
\centering
\fbox{\includegraphics[width=\linewidth]{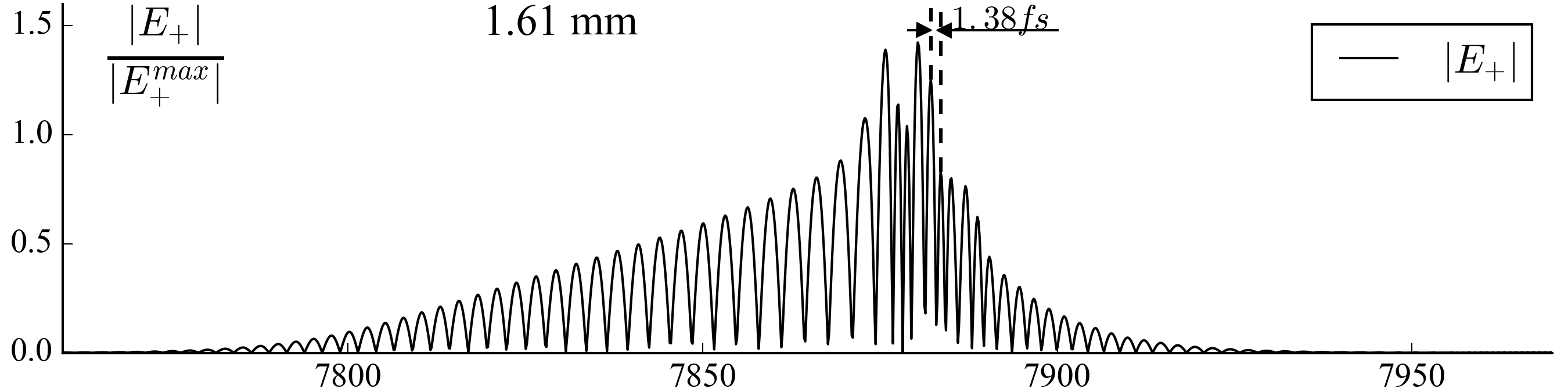}}
\caption{Density of the forward wave propagated over the distance of $1.61\ \textrm{mm}$ with 
extremely steepened tail.}
\label{fig:anomalous_steepened}
\end{figure}

It is well known that a pulse that propagates under conditions of anomalous dispersion and cubic 
nonlinearity experiences strong steepening \cite{trippenbach:98}. For few-cycle pulses the duration 
of the tail can become even shorter than one field cycle \cite{shpolyanskiy:03}. The 
adequacy of unidirectional approximation for such highly nonlinear scenario is questionable, because
an area appears of extremely localized light field. In order to investigate possible features of the
field evolution introduced by inclusion of the backward wave into consideration we simulated the 
propagation of the pulse with central wavelength of $1500\ \textrm{nm}$, duration of 
$50\ \textrm{fs}$ and intensity of $5 \times 10^{9}\ \textrm{kW}/\textrm{cm}^{2}$. Initial 
distribution of the backward wave was, as before, matched with the nonlinear response of the medium 
according to (\ref{eq:backward_initial}). The calculated fields and spectra of the forward and 
backward waves at distances $z$ of $0.0,\ 1.25,\ 1.5$ and $1.85\ \textrm{mm}$ are given in Fig. 
\ref{fig:anomalous_slices_field} and \ref{fig:anomalous_slices_spectrum}, respectively. As shown in 
Fig. \ref{fig:anomalous_steepened}, by the distance of $1.6\ \textrm{mm}$ the tail of the forward 
pulse is considerably steepened. Extreme spectral broadening with splitting into lower and 
high-frequency parts is observed, as is typical for the anomalous dispersion \cite{trippenbach:98}. 
Generation of intense third harmonics is clearly seen in Fig. \ref{fig:anomalous_slices_spectrum}. 
The width of spectral supercontinuum exceeds 1.5 octaves. The pulse tail is formed from 
higher-frequency components and includes features varying sharply in the time scale of a period of 
the initial central frequency (see Fig. \ref{fig:anomalous_steepened}). The forward-propagating part
of the backward wave is again fully defined by the forward pulse and related to it via the analytic 
expression (\ref{eq:backward_simplified_right}). Its spectrum reaches the fifth harmonic. However, 
even this regime does not lead to generation of the reflected wave as can be seen from Fig. 
\ref{fig:anomalous_slices_spectrum}: $G_{-}^{\textrm{left}}$ does not vary from 
$z = 1.25\ \textrm{mm}$ to $z = 1.85\ \textrm{mm}$. The amplitude of $E_{-}^{\textrm{left}}$ is 
about 82 times smaller than the amplitude of $E_{-}^{\textrm{right}}$, which in turn is 
$\sim 0.8 \times 10^{-3}$ of $E_{+}$.

\section{Conclusion}

We have simulated forward and backward waves of few-cycle pulses propagating in single-mode regime 
in telecommunication-type fused-silica fiber using the $z$-propagated approach. Scenarios of pulse 
evolution at intensities of the order of $10^{10}\ \textrm{kW}/\textrm{cm}^{2}$ are considered for 
the normal and anomalous dispersion range of the fiber. Under the combined action of the cubic 
nonlinearity and dispersion the forward wave experiences tremendous changes with generation of 
octave-spanning spectrum. In time domain, a sharp steepening of the pulse tail is observed in 
anomalous-dispersion regime.

The structure of the field which is ignored in the unidirectional approximation is 
investigated analytically and numerically. It includes two parts: one part propagates in the forward
direction along with the forward wave, the other one propagates backwards. It is shown that the 
former can be calculated from the forward wave via the simple expression 
(\ref{eq:backward_simplified_right}) instead of numerical solution of the equation for the backward 
wave. If the matched initial distribution (\ref{eq:backward_initial}) is applied, 
the field propagating backwards becomes considerably weaker than the forward-running part. We 
believe that it is not zero since we ignored dispersion effects in derivation of the matching 
condition (\ref{eq:backward_initial}).

In our numerical experiments, the backward wave remains weak and does not affect the forward wave. It
validates the usage of the unidirectional approximation for such setups at intensities up to the 
order of $10^{10}\ \textrm{kW}/\textrm{cm}^{2}$. The estimate of the amplitude of the backward wave 
given by (\ref{eq:estimate}) is confirmed by numerical results and can be used for a priori 
justification of the application of the unidirectional approach. In the considered scenarios, this 
amplitude was less than $3 \times 10^{-3}$ of the amplitude of the forward wave.

Starting this work, we were going to reveal disadvantages of the unidirectional approximation. But 
our investigations only confirm that it is a brilliant approach for the nonlinear optics in 
transparent media with non-resonant dispersion and nonlinearity. The main drawback is the inability 
to describe injection of a given field profile into the waveguide which is immanent to 
$z$-propagated equations. Neglect of the backward wave does not lead to wrong results for the 
forward wave, but simplifies equations significantly. Instead of the second-order wave equation 
(\ref{eq:wave_equation}) or equivalent system of two coupled equations 
(\ref{eq:direcional_fields_set}) one gets a single first-order equation. It is solved naturally in a
frame running with the group velocity of the forward pulse, which reduces the necessary time window 
dramatically and provides a huge economy of computational resources in numerical simulations.

\begin{acknowledgments}

We are grateful to Ildar Khalidov and Alexander Konev for the proof-reading and stylistic editing.

\end{acknowledgments}

\bibliography{konev_shpolyanskiy_2016}

\end{document}